\documentclass{article}
\usepackage{psfig}
\usepackage{epsf}
\usepackage{emulateapj}
\usepackage{mathptm}
\begin{document}

\def\dd{\partial}
\lefthead{Hawley et al.}
\righthead{Nonradiative Accretion Flow in Three Dimensions}
\slugcomment{Submitted to ApJ Letters March 29, 2001}

\title{A Magnetohydrodynamic Nonradiative Accretion 
Flow in Three Dimensions}

\author{John F. Hawley,\altaffilmark{1} 
Steven A. Balbus\altaffilmark{1},
and James M. Stone\altaffilmark{2}}

\altaffiltext{1}{Dept. of Astronomy, University of Virginia, PO Box
3818, Charlottesville,\\ VA 22903, USA. jh8h@virginia.edu;
sb@virginia.edu}
\altaffiltext{2}{Dept. of Astronomy, University of Maryland,
College Park, MD 20742,\\ USA. jstone@astro.umd.edu}

\begin{abstract}

We present a global magnetohydrodynamic (MHD) three dimensional
simulation of a nonradiative accretion flow originating in a pressure
supported torus.  The evolution is controlled by the magnetorotational
instability which produces turbulence.  The flow forms a nearly
Keplerian disk.  The total pressure scale height in this disk is
comparable to the vertical size of the initial torus.  Gas pressure
dominates only near the equator; magnetic pressure is more important in
the surrounding atmosphere.  A magnetically dominated bound outflow is
driven from the disk.  The accretion rate through the disk exceeds the
final rate into the hole, and a hot torus forms inside 10 $r_g$.  Hot
gas, pushed up against the centrifugal barrier and confined by magnetic
pressure, is ejected in a narrow, unbound, conical outflow.  The
dynamics are controlled by magnetic turbulence, not thermal convection,
and  a hydrodynamic $\alpha$ model is inadequate to describe the flow.
The limitations of two dimensional MHD simulations are also discussed.

\end{abstract}


\section{Introduction}

Observations of underluminous accretion-powered X-ray sources pose a
difficulty for accretion theory.  The Galactic center is a particularly
compelling example:  despite the inferred presence of a 2 million solar
mass black hole, and the availability of a substantial gas reservoir,
only a small fraction of the expected radiation is detected.  In a
standard Keplerian accretion disk, liberated mechanical energy is
promptly radiated, and there is a direct relationship between the
accretion rate $\dot M$ and the luminosity $L$.  A number of
alternatives to the standard disk have been put forward, which are
classified by the ultimate repository of the orbital energy.  In
Advection Dominated Accretion Flows (ADAFs, e.g., Narayan \& Yi 1994),
the energy is retained by the plasma and advected into the hole.  In
contrast, Blandford \& Begelman (1999) proposed that the bulk of the
accreting gas and energy might be carried off by a wind.  A more recent
variation, referred to as Convective Dominated Accretion Flow (CDAF;
e.g., Quataert \& Gruzinov 2000), brings enhanced viscous heating to
the fore, leading to a strong convective flow that stifles the
accretion.  Since the common feature of all these models is negligible
radiative losses, we shall refer to them generically as Non-Radiative
Accretion Flows (NRAFs).

Earlier simulations of NRAFs have used several simplifying
approximations.  Most (e.g., Igumenshchev \& Abramowicz 1999, 2000,
hereafter IA99, IA00; Stone, Pringle, \& Begelman 1999, hereafter SPB)
have assumed the flow evolves due to a fortified kinematic viscosity,
$\nu$.  While illuminating some aspects of NRAFs, there are substantial
drawbacks to this approach.  Most importantly, the results strongly
depend upon the specific recipe adopted for $\nu$.  The origin of
angular momentum transport is not unknown, however.  It arises from the
magnetorotational instability (MRI; Balbus \& Hawley 1991), and
hydrodynamic and magnetohydrodynamic (MHD) flows have fundamentally
different properties.

Stone \& Pringle (2001; hereafter SP) include MHD in their
axisymmetric simulations.  The stress required to drive the accretion
flow emerges self-consistently;  it is not a free parameter.  But the
two-dimensional (2D) restriction is a significant limitation.  First,
the anti-dynamo theorem (e.g., Moffat 1978) prevents the indefinite
maintenance of the poloidal magnetic field in the face of dissipation.
Second, axisymmetric simulations 
tend to over-emphasize the ``channel" mode (Hawley \& Balbus 1992)
which produces coherent internal magnetized flows rather than the more
generic MHD turbulence.  Consequently, a fully self-consistent NRAF
model requires three-dimensional (3D) MHD.  In this {\it Letter} we
present and discuss the results of a prototype 3D NRAF simulation.

\section{The Simulation}

The simulation evolves the equations of ideal MHD, i.e., the
continuity equation, the induction equation, the Euler equation,
\begin{equation}\label{mom}
\rho {\partial{\bf v} \over \partial t}
+ (\rho {\bf v}\cdot\nabla){\bf v} = -\nabla\left(
P + {\mathcal Q} +{B^2\over 8 \pi} \right)-\rho \nabla \Phi +
\left( {{\bf B}\over 4\pi}\cdot \nabla\right){\bf B},
\end{equation}
and an internal energy equation
\begin{equation}\label{ene}
{\partial\rho\epsilon\over \partial t} + \nabla\cdot (\rho\epsilon
{\bf v}) = -(P+{\mathcal Q}) \nabla \cdot {\bf v}.
\end{equation}
The variables have their usual meanings; 
${\mathcal Q}$ is an explicit artificial viscosity (Stone \& Norman
1992a), and $\Phi = - GM/(r-r_g)$ is the
Paczy\'nski \& Wiita (1980) pseudo-Newtonian potential.
We use units with $GM=r_g=1$.  The equation of state is adiabatic,
$P=\rho\epsilon(\Gamma -1)$, with $\Gamma=5/3$.  Radiation transport
and losses are omitted.  Since there is no explicit resistivity or
shear viscosity, the gas can heat only by adiabatic
compression, or by the artificial viscosity $\mathcal Q$.  We 
evolve the equations using
time-explicit Eulerian finite differencing with the ZEUS algorithms
(Stone \& Norman 1992a; 1992b; Hawley \& Stone 1995; Hawley 2000).

The major difficulties of a 3D simulation include the large number of
grid zones needed to resolve an extended spatial domain, and the number
of time steps required to evolve the problem over several orbital
periods at large radius.  In this simulation, we use a somewhat
restricted resolution of $128\times32\times 128$ grid zones in
cylindrical coordinates $(R,\phi,Z)$.  The radial grid extends from
$R=1.5$ to $R=170$.  There are 36 equally-spaced zones inside $R=15$,
and 92 zones outside this point that increase logarithmically in size.
Similarly, 50 $Z$ zones are equally spaced between $-10$ and $10$ with
the remainder of the zones logarithmically stretched to the $Z$
boundaries at $\pm 60$.  The azimuthal domain is limited to $\pi/2$.
Hawley (2001) found that reduced angular coverage preserves qualitative
features of a simulation with only a small reduction in magnetic energy
and stress.  The radial and vertical boundary conditions are simple
zero-gradient outflow conditions; no flow into the computational domain
is permitted.  The $\phi$ boundary is periodic.  The magnetic field
boundary condition is set by requiring the transverse components of the
field to be zero outside the computational domain, while the
perpendicular component satisfies the divergence-free constraint.

The prototype simulation starts with of a constant specific angular
momentum ($\ell$) torus with a pressure maximum at $r=100$ (similar to
model F of SP) and an inner edge at $r=75$.  The initial magnetic field
consists of poloidal loops lying along isodensity contours with a
volume-averaged energy of $\beta = P_{gas}/P_{mag} = 200$.  The average
specific energy of the magnetofluid in the torus is $-4.6\times
10^{-3}$ (in units where $GM=r_g=1$), i.e., it is bound.  The
simulation is run out to 5 orbits at the initial pressure maximum.
This corresponds to 1429 orbits at the marginally stable orbit, $R=3$.
The simulation requires over 1.7 million timesteps with an average
$\Delta t =0.0152$.

\section{Results}

During the first orbit the field in the torus is amplified by the MRI
and by shear, causing the torus to expand.  The MRI acts most
visibly near the equatorial plane, where the field is predominantly
vertical, and the long-wavelength, nearly axisymmetric modes of the MRI
grow rapidly.  The linear growth phase ends shortly after one orbit,
and the magnetic energy has increased to $\beta \approx 2$--10.

Over the next orbit, low-$m$ spiral arms of gas accrete from the inner
edge of the torus, forming a vertically thin and very nonuniform
disk.  Strong magnetic fields surround the gas.  Due to the
initial field topology,  a current sheet forms near the equator which
proves unstable to vertical oscillations.  As more gas leaves the
torus, the disk fills out and thickens.  Low density material is driven
off the forming disk to create a backflow.

As time advances, the initial constant-$\ell$ torus evolves to a nearly
Keplerian angular momentum distribution.  There is net accretion inside
of $R=100$, and net outflow beyond this point.  Inside of $R=100$ the
NRAF forms a modestly thick, nearly Keplerian disk.  In the disk,
accretion results from MHD turbulence, but the final accretion rate
into the black hole need not match the supply rate
precisely, and it does not.  Because the inflow is hot,
geometrically thick, and nearly Keplerian,
inflow is difficult past the marginally
stable orbit.  To accrete the gas must pass through a narrow gap in the
centrifugal barrier at the equator.  Gas splashes off the centrifugal
barrier near the hole, forming a pressure-supported torus and creating a
backflow that adds to a growing low density envelope around a higher
density Keplerian core.

Between orbits 4 and 5, the disk is well-formed and reasonably steady.
Figure 1 is a series of $(R,z)$ contour plots of azimuthally averaged
values at $t=5$ orbits showing that most of the gas is located in a
modestly thick disk surrounded by a low density, highly magnetized
atmosphere.  Near the equator gas pressure is dominant, i.e., $\beta >
1$, but in the surrounding region $\beta < 1$.  The gas pressure scale
height is $H\approx 7$--10, decreasing rapidly inside $R=10$.  The
total pressure (gas plus magnetic) is much smoother, and has a scale
height $H\approx 20$ at $R=100$ that decreases slowly inward.  The gas
density, vertically averaged over one gas pressure scale height, is
nearly constant from $R=50$ down to $R=10$, where it increases rapidly
to a peak at $R=5.5$.  Inside $R=10$, the disk resembles a pressure
supported torus.

The specific angular momentum $l(R)$ tracks a Keplerian distribution
throughout the disk, although in places it is up to 10\% below the
Keplerian value.  The angular velocity $\Omega$ is constant along
cylinders through the main portion of the disk, consistent with
$P=P(\rho)$ there.  In hydrodynamic simulations, surfaces of constant
entropy $S$ coincide with those of specific angular momentum $\ell$.
This corresponds to marginal stability by one of the H\o iland
criteria.  As already noted by SP, with MHD the level surfaces of $S$
and $\ell$ do not line up, particularly inside the main disk.

Magnetic fields in the disk produce Maxwell stress, MHD turbulence, and
angular momentum transport.  Between orbit 4 and 5 the ratio of the
vertical- and $\phi$-averaged Maxwell stress to the gas pressure ranges
from $0.1$ to 0.2 inside of $R=100$.

The mass inflow and outflow rates through every cylindrical radius are
computed as a function of time.  An average over the last orbit shows
that these two are nearly equal, although inflow exceeds outflow (Fig.
2).  Of course the instantaneous inflow and outflow rates merely
represent the flow produced by the MHD turbulent velocity
fluctuations.   The net accretion rate, $\dot M$, is due to the
resulting drift velocity, which is always considerably smaller.

At the end of the run the total mass on the grid has decreased by
5.4\%.  Roughly 51\% of this leaves through the outer radial boundary,
43\% through the upper and lower $Z$ boundaries, and the remaining 6\%
is accreted into the black hole.  A higher resolution torus simulation
(Hawley \& Krolik 2001) shows that the magnetic stress can remain large
down to and beyond the marginally stable orbit.  This effect could increase
$\dot M$, but the present simulation is not sufficiently well-resolved to
address this point.  The flow through the outer radial boundary is a
consequence of the gain in $\ell$ in the outer part of the torus.  The
flow through the $Z$ boundaries is driven from the accretion disk by
gas and magnetic pressure.  Part of this is an unbound, high
temperature hollow conical outflow confined to the axis region by
surrounding magnetic pressure.  This can be seen in the gas pressure
plot of Figure 1.  The remainder of the outflow remains bound, although
its specific energy is greater than that of the initial torus.

The initial average specific energy for the gas is $-4.6\times
10^{-3}$.  At 5 orbits this has decreased to $-4.9\times 10^{-3}$ as
the gas remaining on the grid becomes more bound.  
The total thermal, magnetic, poloidal kinetic, and
orbital energies of the gas that remains on the grid 
all increase with time.  The magnetic energy has
increased the most, receiving 42\% of the total net energy increase;
95\% of this is in toroidal field.  The thermal, orbital and kinetic
energies comprise 25\%, 21\%, and 12\% of the increase respectively.

\section{Discussion}

\subsection{Comparison with $\alpha$ models}

Full 3D MHD simulations are difficult and expensive; are they
necessary?  With respect to the hydrodynamical alternative
the answer is clearly yes.  Such simulations (IA99; SPB; IA00) have
shown that the results depend strongly on the magnitude of $\alpha$ and
the form assumed for the stress.   Thus, a self-consistent stress is
essential for understanding NRAF dynamics.

One should not lose sight of the fact that this is a high Reynolds
number turbulent system, not a low Reynolds number laminar flow.  For
example, the result that large Shakura-Sunyaev $\alpha$ ($\nu=\alpha
c_s^2/\Omega$) accretion flows are stable, laminar and accrete fully
into the central hole (IA99; IA00) is a literal consequence of using a
Navier-Stokes viscosity for the stress.  With stress from MHD
turbulence $\alpha \sim 1$ necessarily implies velocity fluctuations
with speeds comparable to the sound speed $c_s$ on scales of order the
size of the system.

A second hydrodynamic result is that NRAFs are driven to a state of
marginally convective instability  (SPB; Quataert \& Gruzinov 2000;
IA00).  This is the basis for CDAF solutions.  But this 
cannot hold in MHD.  Recall the form of the H\o iland criteria
that applies to adiabatic perturbations in {\it magnetized} flow
(Balbus 1995):
\begin{equation}
N^2 + {\dd \Omega^2\over \dd \ln R} > 0
\end{equation}
\begin{equation}
- \left(\dd P\over \dd Z\right)\left( {\dd\Omega^2\over\dd R} {\dd
\ln P\rho^{-5/3}\over \dd Z} -  {\dd\Omega^2\over\dd Z} {\dd
\ln P\rho^{-5/3}\over \dd R} \right) > 0
\end{equation}
where $N^2 = -(3/5\rho) \nabla P \cdot \nabla\ln P\rho^{-5/3}$
is the Brunt-V\"ais\"al\"a frequency.
Note the symmetry between the angular velocity gradients and
the entropy gradients, and the absence of any stabilizing epicyclic
frequency.  At least one of these criteria must always be strongly, not
marginally, violated in an accretion flow:  this is what is responsible
for the existence of the turbulent stress.  Convection does not arise
{\it in addition} to this fundamental instability, it is part of the
instability that produces the anomalous stress in the first place.  In
particular, one cannot argue for the marginal stability of 
a rotationally modified convective process independently of
what is called $\alpha$.  Whatever the sign of the Brunt-V\"ais\"al\"a
frequency, the instability is essentially the same one at work in a
standard Keplerian disk.  Indeed, in contrast to hydrodynamical
treatments, we find that the MHD flows show no tendency at all toward
marginal stability. 

The 3D MHD simulation does show large scale turbulent rolls and
outflows qualitatively similar to those reported in small $\alpha$
hydrodynamic simulations.  In the hydrodynamic simulations, however,
all the rotational energy released by the stress goes immediately into
heat.  Some of this heat returns to kinetic energy in the form of
large-scale convection.   This implies that large-scale shear is
converted into heat, producing large-scale convective rolls, which then
prevent the accretion process that was driving the heating in the first
place.  In MHD, the orbital energy goes directly into turbulent kinetic
and magnetic energies.  The magnetic energy constitutes a particularly
important component of the total energy budget.  Heating results from
dissipation at small scales.  In a real plasma, this will depend on
microscopic resistivity and viscosity.  In the simulation, turbulent
rolls are not thermal in origin.   More generally the kinetic energy of
convective motions is small compared with the dynamical activity
associated with MHD instabilities and with the large scale flow
itself.

We also remark that in any turbulent flow, the net accretion rate will
always be small compared to any locally sampled instantaneous value.
As Figure 2 shows, inward and outward accretion rates almost exactly
cancel in our purely dynamical simulation.  This is not a
unique property of inwardly-transported convective energy fluxes.  It
is a hallmark of turbulence.  In both viscous hydrodynamical and MHD
treatments, turbulent eddy velocities are much larger than the residual
inward drift velocity.  It is precisely this ordering that allows the
$\alpha$ formalism for the {\it mean} flow dynamics to be derived from
the equations of motion themselves (Balbus \& Papaloizou 1999).

\subsection{Comparison with 2D MHD models}

Next we compare the present simulation with the 2D models of SP.  In
the equivalent run (SP's Run F), the initial infall phase during orbits
2--3 was relatively smooth and dominated by the channel flow of the
MRI.  Subsequent evolution was turbulent.  In 3D, the initial smooth
phase does not occur; the evolution is always turbulent.
The 3D flow also generates greater outflow from the disk.

After 3 orbits in both 2D and 3D, the inner regions of the flow are
dominated by MHD turbulence driven by the MRI.  Interestingly, there
appear to be few differences between time- and angle-averaged
quantities in either case.  For example, comparison of the contours in
Fig.~1 with Figure 4 in SP show that both produce a nearly barytropic
disk near the midplane with $\Omega$ contours parallel to cylindrical
radii, but with no correlation between the specific angular momentum
and entropy.  Moreover, the net $\dot M$ values are similar: $3\times
10^{-3}$ in 2D versus $\sim 2 \times 10^{-3}$ in 3D, in units of the
initial mass of the torus per orbit at the initial pressure maximum.
Radial profiles of time-averaged data in the main disk show many
similarities: in both cases $P_{gas} \sim 10 P_{mag}$ and $v_{\phi} >
c_{s} > v_{Alfven}$ at all radii (see figure 6 in SP).  The most
obvious differences between the 2D and 3D models are near the axis, and
these are attributable to spherical 2D versus cylindrical 3D inner
boundary geometries.

Eventually all axisymmetric models are limited by the antidynamo
theorem.  Toward the end of the SP simulations the turbulence in the
initial torus dies down, and remains only in the flow close to the
black hole.  Higher resolution is possible in 2D, but this only
prolongs the duration of the turbulent phase before eventual decline.
In addition, the toroidal field MRI cannot be simulated in axisymmetry,
which limits 2D models to initial conditions that contain large-scale
poloidal field.

\section{Conclusion}

We have performed a three-dimensional MHD simulation of an NRAF
originating from a constant-$\ell$ torus located at 100 gravitational
radii.  The resulting flow does not resemble a classical ADAF.  Most of
the infalling mass and released energy does not end in the hole.   In
this sense the flow might be better described as an inflow/outflow
solution of the type outlined by Blandford \& Begelman (1999).  The
general appearance of the flow is similar to the low-$\alpha$
hydrodynamic models of SPB, IA99 and IA00.  Of course this is what one
should expect for any hot, rotating, nearly Keplerian accretion flow.
We have argued that the CDAF picture is also inappropriate because the
the classic H\o illand criteria do not apply to an MHD flow.   The
outflow is not driven by thermal convection, but by the action of the
same instability that accounts for $\alpha$, namely the MRI.
Marginally stability is not attained.

We conclude by considering some directions for future work.  In the
present simulation an internal energy equation is used and the only
source of dissipative heating is the artificial viscosity ${\mathcal
Q}$.  Some amount of the total energy is inevitably lost in this
formulation.  A portion of this can be recaptured through artificial
resistivity, although this made no significant difference in flow
structure or accretion rates when tested in 2D (SP).  In any case, a
more detailed study of the energy flow in a 3D simulation would be
enlightening.  Higher resolution is needed in the inner region to
capture the details of the inner disk boundary and the flow past the
marginally stable orbit.  Models with a greater radial domain would
allow the inflow to proceed over several decades in distance and to
liberate greater amounts of gravitational energy.  High-resolution
axisymmetric MHD simulations can play an important role here in mapping
out the types of flows that are possible.  Longer evolutions would
provide greater confidence that the results have lost memory of the
initial conditions, but, because of the antidynamo theorem, true steady
states can only be obtained in 3D.

\acknowledgements{JFH and SAB acknowledge support under NSF grant
AST-0070979, and NASA grants NAG5-9266, NAG5-7500, and NAG5-106555.  
JMS acknowledges support under NASA grant NAG5-3836.
The simulation
was carried out on the IBM Bluehorizon system of the San Diego
Supercomputer Center of the National Partnership for Advanced
Computational Infrastructure, funded by the NSF.  }

% figure 1
\begin{figure*}
\centerline{
\psfig{figure=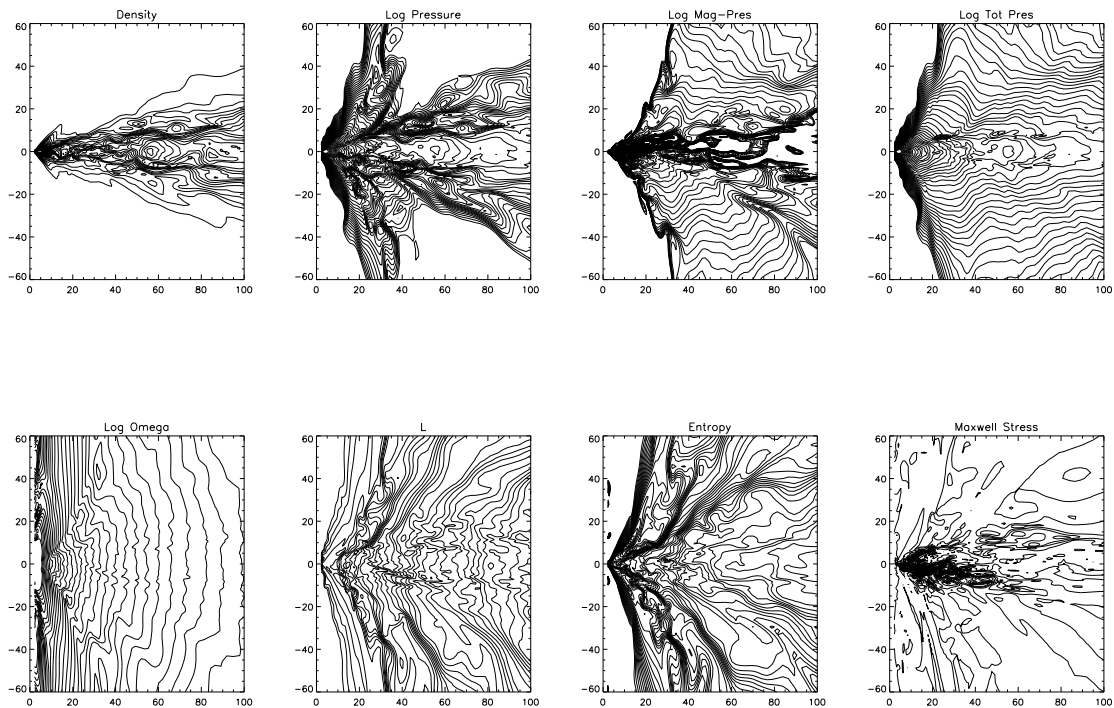,width=6.0in,angle=90}}
\caption{Azimuthally-averaged contour plots of indicated
quantities in the inner region of the computation at $t=5$ orbits.  
Plots are gas density, log of the gas pressure, log of the magnetic
pressure, log of the total pressure, log of the angular velocity
$\Omega$, specific angular momentum $\ell$, the entropy, $S =\ln
(P/\rho^{5/3})$, and the Maxwell stress, $T_{R\phi} =-B_R B_\phi/4\pi$.
}
\end{figure*}

% figure 2
\begin{figure*}
\centerline{
\psfig{figure=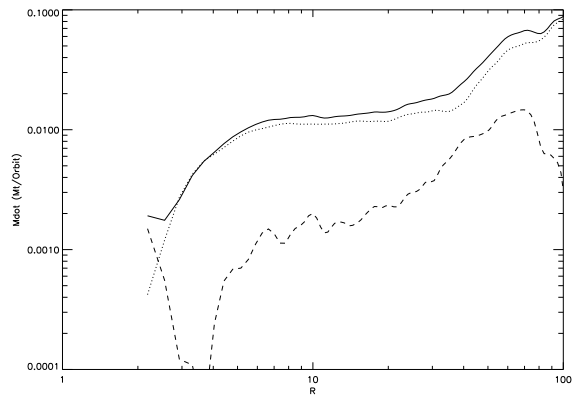,width=3.0in,angle=90}}
\caption{Amplitude of the total mass inflow rate (solid
line), mass outflow rate (dotted line), and net accretion rate (dashed
line) through cylindrical radius $R$, averaged in time
between orbit 4 and 5. }
\end{figure*}

\end{document}